\documentclass[letterpaper, 10pt, conference]{ieeeconf}

\IEEEoverridecommandlockouts                              
\usepackage{graphicx}
\usepackage[dvips]{epsfig}

\usepackage{amssymb,amsmath,amsfonts,enumerate,bbm}
\usepackage{cite}
\usepackage{caption,subcaption}
\usepackage{epstopdf}

\usepackage{csquotes}
\usepackage[UKenglish,USenglish]{babel}

\def\la{\lambda}
\def\La{\Lambda}
\def\vp{\varphi}
\def\ve{\varepsilon}

\def\g{\mathcal G}

\def\r{\mathbb R}

\def\be{\beta}

\def\ones{\mathbbm{1}}

\newcommand{\dfb}{\stackrel{\Delta}{=}}

\def\be{\begin{equation}}
\def\ee{\end{equation}}
\def\ben{\begin{equation*}}
\def\een{\end{equation*}}

\newtheorem{thm}{Theorem}
\newtheorem{lem}{Lemma}
\newtheorem{cor}{Corollary}
\newtheorem{defn}{Definition}
\newtheorem{rem}{Remark}
\newtheorem{assum}{Assumption}
\newtheorem{prop}{Proposition}

\title{Modulus consensus in discrete-time signed networks and properties of special recurrent inequalities}

\author{Anton V. Proskurnikov and Ming Cao%
\thanks{A.V. Proskurnikov is with the Delft Center for Systems and Control (DCSC) at Delft University of Technology. 
He is also with ITMO University and Institute for Problems of Mechanical Engineering of Russian Academy of Sciences (IPME RAS), St. Petersburg, Russia;\;{\tt\small anton.p.1982@ieee.org}}%
\thanks{M. Cao is with the Engineering and Technology Institute (ENTEG) at the University of Groningen, The Netherlands;\;{\tt\small m.cao@rug.nl}}%
\thanks{Financial support was provided by the ERC (ERC-StG-307207), NWO (vidi-438730),
Russian Federation President's Grant MD-6325.2016.8, and RFBR, grants 17-08-01728, 17-08-00715 and 17-08-01266. 
Theorem~1 is obtained at IPME RAS under the sole support of Russian Science Foundation (RSF) grant 14-29-00142.}%
}

\begin{document}

\maketitle
\thispagestyle{empty}
\pagestyle{empty}

\begin{abstract}
Recently the dynamics of signed networks, where the ties among the agents can be both positive (attractive) or negative (repulsive)
have attracted substantial attention of the research community. Examples of such networks are models of opinion dynamics over signed graphs, 
recently introduced by Altafini (2012,2013) and extended to discrete-time case by Meng et al. (2014). It has been shown that under 
mild connectivity assumptions these protocols provide the convergence of opinions in absolute value, whereas their signs may differ.
This ``modulus consensus'' may correspond to the \emph{polarization} of the opinions (or \emph{bipartite consensus}, including the usual consensus as a special case), or their convergence to zero.

In this paper, we demonstrate that the phenomenon of modulus consensus in the discrete-time Altafini model is a manifestation of a more
general and profound fact, regarding the solutions of a special recurrent inequality. Although such a recurrent inequality does not provide the uniqueness of a solution,
it can be shown that, under some natural assumptions, each of its bounded solutions has a limit and, moreover, converges to consensus.
A similar property has previously been established for special continuous-time differential inequalities (Proskurnikov, Cao, 2016).
Besides analysis of signed networks, we link the consensus properties of recurrent inequalities to the convergence analysis of distributed optimization algorithms and
the problems of Schur stability of substochastic matrices.
\end{abstract}

\section{Introduction}

In the recent years protocols for consensus and synchronization in multi-agent networks have been thoroughly studied~\cite{Murray:07,RenBeardBook,RenCaoBook,LewisBook}. Much less studied are ``irregular'' behaviors,
exhibited by many real-world networks, such as e.g. desynchronization~\cite{Pecora:14} and chaos~\cite{Popovych:05}.
An important step in understanding these complex behaviors is to elaborate mathematical models for ``partial'' or cluster synchronization, or simply \emph{clustering}~\cite{Pecora:14,Pogromsky08,XiaCao:11,ProCao16-3}. In social influence theory,
this problem is known as the \emph{community cleavage} problem or Abelson's \emph{diversity puzzle}
~\cite{Friedkin:2015,ProTempo:2017-1}: to disclose mechanisms that hinder reaching consensus among social actors
and lead to splitting of their opinions into several clusters.

One reason for clustering in multi-agent networks is the presence of ``negative'' (repulsive, antagonistic) interactions among the agents~\cite{XiaCao:11}. Models of \emph{signed} (or ``coopetition'') networks with positive and negative couplings among the nodes describe a broad class of real-world systems, from molecular ensembles~\cite{LinLeeFuhJuanHuang:13} to continental supply chains~\cite{Pache:13}. Positive and negative relations among social actors can express, respectively, trust (friendship) or distrust (hostility). Negative ties among the individuals may also result from the \emph{reactance} or \emph{boomerang} effects, first described in \cite{HovlandBook}: an individual may not only resist the persuasion process, but even adopt
an attitude that is contrary to the persuader's one.

A simple yet instructive model of continuous-time opinion dynamics over signed networks has been proposed by Altafini~\cite{Altafini:2012,Altafini:2013} and extended to the discrete-time case in~\cite{MengCaoJohansson:2014}. In the recent years, Altafini-type coordination protocols over static and time-varying signed graphs have been extensively studied, see e.g.
~\cite{Hendrickx:14,LiuChenBasar:15,XiaCaoJohansson:16,ProMatvCao:2016,Meng:16-2,MengShiCao:16,LiuChenBasar:16,LiuChamieBasarAcikmese:16}.
It has been shown that under mild connectivity assumptions these models exhibit consensus in absolute value, or \emph{modulus consensus}: the agents' opinions agree in modulus yet may differ in signs.
The modulus consensus may correspond to the asymptotic stability of the network (the opinions of all individuals converge to zero), usual consensus (convergence of the opinions to the same value,
depending on the initial condition) and \emph{polarization}, or ``bipartite consensus'': the agents split into two groups, converging to the opposite opinions.

In the recent works~\cite{ProCao16-4,ProCao:2017} it has been shown that the effect of modulus consensus in the continuous-time Altafini model is in fact a manifestation of a more profound result, concerned with the special class of \emph{differential inequalities}
\be\label{eq.ineq0}
\dot x(t)\le -L(t)x(t),
\ee
where $L(t)$ stands for the Laplacian matrix of a time-varying weighted graph. Although the inequality~\eqref{eq.ineq0} 
is a seemingly ``loose'' constraint, any of its \emph{bounded} solutions (under natural connectivity assumptions) converges to a consensus equilibrium (this property is called \emph{consensus dichotomy}). This implies, in particular, the modulus consensus in the Altafini model~\cite{ProCao16-4,ProCao:2017} since the vector of the opinions' absolute values obeys the inequality~\eqref{eq.ineq0}.
In this paper, we extend the theory of differential inequalities to the discrete-time case, where~\eqref{eq.ineq0} is replaced by the \emph{recurrent} inequality $x(k+1)\le W(k)x(k)$ with $\{W(k)\}_{k\ge 0}$ being a sequence of stochastic matrices. We establish the consensus dichotomy criteria for these inequalities, which imply the recent results on modulus consensus in the discrete-time Altafini model~\cite{MengCaoJohansson:2014,MengShiCao:16}. We also apply the recurrent inequalities to some problems of matrix theory and the analysis of distributed algorithms for optimization and linear equations solving. 
\section{Problem Setup}

We start with preliminaries and introducing some notation.

\subsection{Preliminaries}

 First we introduce some notation. A vector $x\in\r^n$ is non-negative ($x\ge 0$) if $x_i\ge 0\,\forall i$. Given two vectors $x,y\in\r^n$, we write $x\ge y$ (respectively, $x\le y$) if $x-y\ge 0$ (respectively, $y-x\ge 0$). The vector of ones is denoted by $\ones_n=(1,\ldots,1)^{\top}\in\r^n$. Given a matrix $A=(a_{ij})$, we use $|A|=(|a_{ij}|)$ to denote the matrix of element-wise absolute values (the same rule applies to vectors). A matrix $A=(a_{ij})$ is \emph{stochastic} if its entries are non-negative and all rows sum to $1$, i.e. $\sum_ja_{ij}=1\,\forall i$.
We use $\rho(A)$ to denote the spectral radius of a square matrix $A$. The standard Euclidean norm of a vector $x$ is denoted by $\|x\|=\sqrt{x^{\top}x}$.

A non-negative matrix $A=(a_{ij})_{i,j\in V}$ can be associated to a (directed) weighted graph\footnote{We assume that the reader is familiar with the standard concepts of graph theory, regarding directed graphs
and their connectivity properties, e.g. walks (or paths), cycles and strongly connected components~\cite{HararyBook:1965,BulloBook-Online}.} $\g[A]=(V,E[A],A)$, whose set of arcs is $E[A]=\{(i,j):a_{ij}\ne 0\}$.

\subsection{Recurrent inequalities and consensus dichotomy.}

In this paper, we are interested in the solutions of the following discrete-time, or \emph{recurrent}, inequality
\be\label{eq.ineq}
x(k+1)\le W(k)x(k),\quad k=0,1,\ldots
\ee
where $x(k)\in\r^n$ is a sequence of vectors and $W(k)\in\r^{n\times n}$ stands for a sequence of \emph{stochastic} matrices.

Replacing the inequality in~\eqref{eq.ineq} by the equality, one obtains the well-known averaging, or \emph{consensus} protocol~\cite{Jad:03,Blondel:05,CaoMorse:08Part1}
\be\label{eq.eq}
x(k+1)=W(k)x(k),
\ee
dating back to the early works on social influence~\cite{French:1956,Harary:1959}, rational decision making~\cite{DeGroot} and distributed optimization~\cite{Tsitsiklis:86}.
The algorithm~\eqref{eq.eq} may be interpreted as the dynamics of opinions\footnote{In the broad sense, ``opinion'' is just a scalar quantity of interest; it can stand for e.g. a physical parameters
or an attitude to some event or issue.} formation in a network of $n$ agents. At each step of the opinion iteration $k$ agent $i$
calculates the weighted average of its own opinion $x_i(k)$ and the others' opinions; this average is used as the new opinion of the $i$th agent $x_i(k+1)=\sum_jw_{ij}(k)x_j(k)$.
The graph $\g[W(k)]$ naturally represents the interaction topology of the network at step $k$. Agent $i$ is influenced by agent $j$ if $w_{ij}(k)>0$, otherwise the $j$th agent's opinion
$x_j(k)$ plays no role in the formation of the new agent $i$'s opinion $x_i(k+1)$.

A similar interpretation can be given to the inequality~\eqref{eq.ineq}. Unlike the algorithm~\eqref{eq.eq}, the opinion of agent $i$ at each step of opinion formation is not uniquely determined by the opinions from the previous step, but is only \emph{constrained} by them $x_i(k+1)\le\sum_jw_{ij}(k)x_j(k)$. The weight $w_{ij}(k)$ stands for the contribution of agent $j$'s opinion $x_j(k)$ to this constraint, and in this sense it can also be treated as the ``influence'' weight. The inequality~\eqref{eq.ineq} does not provide the solution's uniqueness for a given $x(0)$, but only guarantees the existence of an \emph{upper bound} for the solutions.
\begin{prop}~\label{prop.triv}
Any solution of~\eqref{eq.ineq} obeys the inequality
\[
x(k)\le M\ones_n,\quad M\dfb\max_ix_i(0).
\]
\end{prop}
\begin{proof}
Proposition~\ref{prop.triv} is proved via straightforward induction on $k$. By definition, $x(0)\le M\ones_n$; if $x(k)\le M\ones_n$ then $x(k+1)\le W(k)x(k)\le MW(k)\ones_n=M\ones_n$.
\end{proof}

Although many solutions of~\eqref{eq.ineq} are unbounded from below, under certain assumptions any its \emph{bounded} solution converges to a consensus equilibrium $c\ones_n$, where $c\in\r$. A similar property, called \emph{consensus dichotomy}\footnote{The term \emph{dichotomy} originates from ODE theory. A system is dichotomic if any of its solutions either grows unbounded or has a finite limit~\cite{Yak:88SCL}.} has been established in~\cite{ProCao16-4,ProCao:2017} for the differential inequalities~\eqref{eq.ineq0}.
\begin{defn}
The inequality~\eqref{eq.ineq} is said to be \emph{dichotomic} if any of its bounded (from below) solutions has a limit $x_*=\lim\limits_{k\to\infty}x(k)$. It is called \emph{consensus dichotomic} if these limits are consensus equilibria $x_*=c_*\ones_n$, where $c_*\in\r$.
\end{defn}

The main goal of this paper is to disclose criteria of consensus dichotomy in the recurrent inequalities~\eqref{eq.ineq}. In Section~\ref{sec.appl} we discuss applications of these criteria
to models of opinion dynamics and algorithms of distributed optimization.

\section{Main Results}\label{sec.main}

The first step is to examine \emph{time-invariant} inequalities~\eqref{eq.ineq}.

\subsection{A dichotomy criterion for the time-invariant case}

\begin{figure}[h]
\begin{subfigure}[b]{0.49\columnwidth}
\center
\includegraphics[width=0.9\columnwidth]{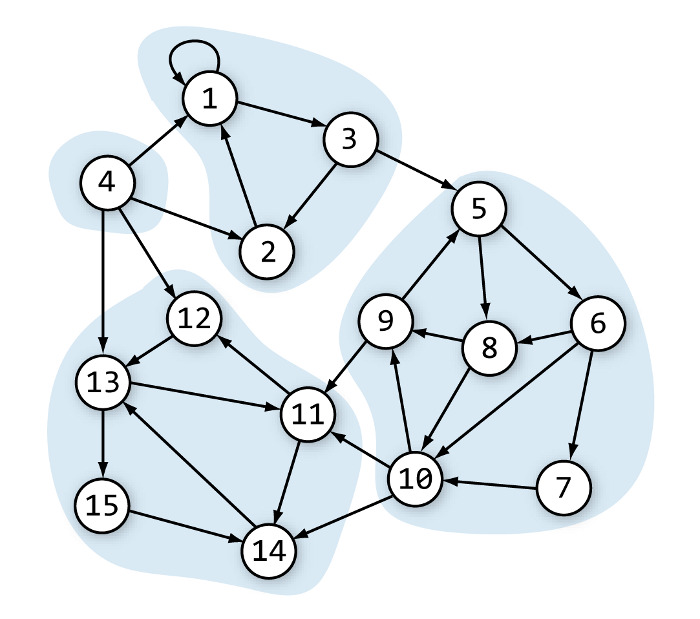}
\caption{}
\end{subfigure}\hfill
\begin{subfigure}[b]{0.49\columnwidth}
\center
\includegraphics[width=0.9\columnwidth]{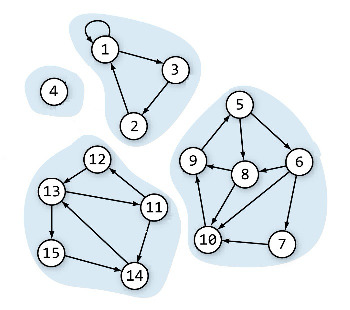}
\caption{}
\end{subfigure}
\caption{Non-isolated (a) vs. isolated (b) strong components}\label{fig.comp}
\end{figure}

In this subsection, we assume that $W(k)\equiv W$ is a constant matrix, whose graph $\g\dfb\g[W]$ has $s\ge 1$ strongly connected (or \emph{strong}) components $\g_1,\ldots,\g_s$; in general,
arcs between different components may exist (Fig.~\ref{fig.comp}a). A
strong component is \emph{isolated} if no arc enters or leaves it. All strong components are isolated (Fig.~\ref{fig.comp}b) if and only if every arc of the graph belongs to a cycle~\cite[Theorem~3.2]{HararyBook:1965}.

\begin{thm}\label{thm.1}
The inequality~\eqref{eq.ineq} with the static matrix $W(k)\equiv W$ is dichotomic if and only if all the strong components $\g_1,\ldots,\g_s$ of its graph $\g$ are isolated and aperiodic\footnote{Recall that a graph is \emph{aperiodic} if the greatest common divisor of its cycles' lengths (that is also referred to as the graph's \emph{period}) is equal to $1$.}.
The inequality is consensus dichotomic if and only if $\g$ is strongly connected ($s=1$) and aperiodic, or, equivalently, the matrix $W$ is \emph{primitive}~\cite{BulloBook-Online,Meyer2000Book}.
\end{thm}

The proof of Theorem~\ref{thm.1}, as well as the remaining results of this section, is given in Appendix.

\begin{rem}
Let $V_j$ stand for the set of nodes of $\g_j$. Theorem~\ref{thm.1} shows that the time-invariant dichotomic inequality~\eqref{eq.ineq} reduces to $s$ independent inequalities of lower dimensions
\be\label{eq.ineq-many}
x^{(m)}(k+1)\le W^{(m)}x^{(m)}(k),\quad m=1,\ldots,s,
\ee
where $x^{(m)}(k)=(x_i(k))_{i\in V_m}$, $W^{(m)}=(w_{ij})_{i,j\in V_m}$ and each inequality~\eqref{eq.ineq-many} is consensus dichotomic.
\end{rem}
\begin{rem}\label{rem.2}
The matrix is primitive if and only if~\cite{BulloBook-Online,Meyer2000Book,ProTempo:2017-1} its powers $W^k$ are strictly positive for large $k$.
\end{rem}

\subsection{Consensus dichotomy in the time-varying case}

In this subsection, we extend the result of Theorem~\ref{thm.1} to the case of general time-varying inequality~\eqref{eq.ineq}.
Given $\ve>0$, let $S_{\ve}$ denote the class of all stochastic matrices $W=(w_{ij})_{i,j\in V}$, satisfying the two conditions:
\begin{enumerate}
\item $w_{ii}\ge\ve$ for any $i\in V$;
\item the graph $\g_{\ve}[W]=(V,E_{\ve}[W])$ is strongly connected, where $E_{\ve}[W]\dfb\{(i,j)\in V\times V:w_{ij}\ge\ve\}$.
\end{enumerate}
In other words, removing from the graph $\g[W]$ all ``light'' arcs weighted by less than $\ve$, the remaining subgraph $\g_{\ve}[W]$ is strongly connected and has self-loops at each node.

For any integers $k\ge 0$ and $m>k$ let $\Phi(m,k)=(\vp_{ij}(m,k))_{i,j=1}^n\dfb W(m-1)\ldots W(k)$ stand for the evolutionary matrix of the equation~\eqref{eq.eq}; for convenience, we denote $\Phi(k,k)=I_n$.
It is obvious that any solution of~\eqref{eq.ineq} satisfies also the family of inequalities
\[
x(m)\le\Phi(m,k)x(k)\quad\forall m\ge k\ge 0.
\]

The following theorem provides a consensus dichotomy criterion for the case of the time-varying matrix $W(k)$.
\begin{thm}\label{thm.2}
The inequality~\eqref{eq.ineq} is consensus dichotomic if $\ve>0$ exists that satisfies the following condition:
for any $k\ge 0$ there exists $m>k$ such that $\Phi(m,k)\in S_{\ve}$.
\end{thm}

Notice that for the static matrix $W(k)\equiv W$ one has $\Phi(m,k)=W^{m-k}$, so the condition from Theorem~\ref{thm.2} means that $W^s\in S_{\ve}$ for some $s$. It can be easily shown that in this case $W^{s(n-1)}$ is a strictly positive matrix. On the other hand, if $W^d$ is strictly positive for some $d$, then $W^d\in S_{\ve}$ for sufficiently small $\ve>0$.
In view of Remark~\ref{rem.2} and~Theorem~\ref{thm.1}, in the static case $W(k)\equiv W$ the sufficient condition of consensus dichotomy from Theorem~\ref{thm.2} is in fact also \emph{necessary},
boiling down to the primitivity of $W$.

The condition from Theorem~2 is implied by the two standard assumptions on the sequence $\{W(k)\}_{k\ge 0}$.
\begin{assum}\label{ass.positive}
There exists $\delta>0$ such that for any $k\ge 0$
\begin{enumerate}
\item $w_{ii}(k)\ge\delta$ for any $i=1,\ldots,n$;
\item for any $i,j$ such that $i\ne j$ one has $w_{ij}(k)\in\{0\}\cup [\delta;1]$.
\end{enumerate}
\end{assum}
\begin{assum}(Repeated joint strong connectivity)\label{ass.uniform}
There exists an integer $B\ge 1$ such that the graph
$\g[W(k)+\ldots+W(k+B-1)]$ is strongly connected for any $k$.
\end{assum}

\begin{cor}\label{cor.1}
Let Assumptions~\ref{ass.positive} and~\ref{ass.uniform} hold. Then the inequality~\eqref{eq.ineq} is consensus dichotomic.
\end{cor}
\begin{proof}
We are going to show that the condition from Theorem~\ref{thm.2} holds for $\ve=\delta^B$ and $m=k+B$, i.e. $\Phi(k+B,k)\in S_{\delta^B}$ for any $k$.
Indeed, $\vp_{ii}(m,k)\ge w_{ii}(m-1)\ldots w_{ii}(k)\ge\delta^{m-k}\,\forall i$ whenever $m\ge k$ due to Assumption~\ref{ass.positive}. Supposing that $(i,j)\in\g[W(l)]$, where $k\le l<m$, one has $\Phi(m,k)=\Phi(m,l+1)W(l)\Phi(l,k)$, and therefore $\vp_{ij}(m,k)\ge \vp_{ii}(m,l+1)w_{ij}(l)\vp_{jj}(l,k)\ge\delta^{m-l-1}\delta\delta^{l-k}=\delta^{m-k}$. Applying this to $m=k+B$, one easily notices that
$i$ is connected to $j$ in the graph $\g_{\delta^B}[\Phi(k+B,k)]$ whenever $w_{ij}(l)>0$ for some $l=k,\ldots,k+B-1$.
 Assumption~\ref{ass.uniform} implies now that $\Phi(k+B,k)\in S_{\delta^B}$ for any $k$.
\end{proof}

It should be noticed however that the condition of Theorem~\ref{thm.2} may hold in many situations where Assumptions~\ref{ass.positive} and~\ref{ass.uniform} fail.
Even in the static case $W(k)\equiv W$, the matrix $W$ can be primitive yet have zero diagonal entries. The following corollary illustrates another situation where both Assumptions~\ref{ass.positive} and~\ref{ass.uniform} may fail, whereas Theorem~\ref{thm.2} guarantees consensus dichotomy.
\begin{cor}\label{cor.2}
Suppose that for any $k$ one has $W(k)\in \{W_0\}\cup\mathcal W$, where $W_0$ stands for the \emph{primitive} matrix and $\mathcal W$ is a set of stochastic matrices, commuting with $W_0$:
$W_0W=WW_0\,\forall W\in\mathcal W$. Let the set $K_0=\{k:W(k)=W_0\}$ be infinite. Then the inequality~\eqref{eq.ineq} is consensus dichotomic.
\end{cor}
\begin{proof}
Let $d$ be so large that $W_0^d$ is a positive matrix, whose minimal entry equals $\ve>0$.
For any $k$, we can find such $m>k$ that the sequence $k,k+1,\ldots,m-1$ contains $d$ elements from the set $K_0$. Since any $W(j)$ commutes with $W_0$, $\Phi(m,k)=T_kW_0^d$, where $T_k$ is some stochastic matrix, and thus all entries of $\Phi(m,k)$ are not less than $\ve$.
\end{proof}

Many sequences $\{W(k)\}$, satisfying the conditions of Corollary~\ref{cor.2}, fail to satisfy Assumptions~\ref{ass.positive} and~\ref{ass.uniform}.
For instance, if $\mathcal W\ni I_n$ then the sequence $\{W(k)\}$ can contain an arbitrary long subsequence of consecutive identity matrices,
which violates Assumption~\ref{ass.uniform}. Both the matrix $W_0$ and matrices from $\mathcal W$ may have zero diagonal entries, which also violates Assumption~\ref{ass.positive}. The set $\mathcal W$ can also be non-compact, containing matrices with arbitrary small yet non-zero entries.

\subsection{The case of bidirectional interaction}

It is known that in the case of bidirectional graphs $w_{ij}>0\Leftrightarrow w_{ji}>0$ the conditions for consensus in the network~\eqref{eq.eq} is reached under very modest connectivity assumptions. Under Assumption~\ref{ass.positive}, consensus is reached if and only if the following relaxed version of Assumption~\ref{ass.uniform} holds~\cite{Blondel:05}.
\begin{assum}(Infinite joint strong connectivity)\label{ass.inf}
The graph $\g_{\infty}=(V,E_{\infty})$ is strongly connected, where
\[
E_{\infty}=\left\{(i,j):\sum_{k=1}^{\infty}w_{ij}(k)=\infty\right\}.
\]
\end{assum}

The following result extends this consensus criterion to the condition of consensus dichotomy in the inequality~\eqref{eq.ineq}.
\begin{thm}\label{thm.3}
Suppose that Assumption~\ref{ass.positive} and~\ref{ass.inf} hold and for any $k$ one has $w_{ij}(k)>0\Leftrightarrow w_{ji}(k)>0$. Then the inequality~\eqref{eq.ineq} is consensus dichotomic.
\end{thm}

The relaxation of Assumption~\ref{ass.positive} in Theorem~\ref{thm.3} remains a non-trivial open problem. To the best of the authors' knowledge, the same applies to usual consensus algorithms~\eqref{eq.eq}: most of the existing results for consensus in discrete-time switching networks~\cite{Blondel:05,Jad:03,CaoMorse:08Part1,RenBeardBook}
rely on Assumption~\ref{ass.positive} or at least require uniformly positive diagonal entries $w_{ii}(k)$.

\section{Examples and Applications}\label{sec.appl}

In this section we apply the criteria from Section~\ref{sec.main} to the analysis of several multi-agent coordination protocols. 

\subsection{Modulus consensus in the discrete-time Altafini model}

We first consider the discrete-time Altafini model~\cite{MengCaoJohansson:2014,LiuChenBasar:15} of opinion formation in a signed network. This model is similar to the consensus protocol~\eqref{eq.eq} and is given by
\be\label{eq.altaf}
\begin{gathered}
\xi(k+1)=A(k)\xi(k)\in\r^n,\quad\text{or, equivalently}\\
\xi_i(k+1)=\sum_{j=1}^na_{ij}(k)x_j(k).
\end{gathered}
\ee
Here the matrix $(a_{ij}(k))$ satisfies the following assumption.
\begin{assum}
For any $k=0,1,\ldots$, the matrix $A(k)=(a_{ij}(k))$ has non-negative diagonal entries $a_{ii}(k)\ge 0$, and the modulus matrix $|A(k)|=(|a_{ij}(k)|)$ is stochastic.
\end{assum}

The non-diagonal entries $a_{ij}(k)$ in~\eqref{eq.altaf} may be both positive and negative. Considering the elements  $\xi_i(k)$ as ``opinions'' of $n$ agents, the positive value
$a_{ij}(k)>0$ can be treated as trust or attraction among agents $i$ and $j$. In this case, agent $i$ shifts its opinion towards the opinion of agent $j$. Similarly, the negative value $a_{ij}(k)<0$ stands for distrust or repulsion among the agents: the $i$th agent's opinion is shifted away from the opinion of agent $j$.
The central question concerned with the model~\eqref{eq.altaf} is reaching consensus in absolute value, or \emph{modulus consensus}~\cite{MengCaoJohansson:2014}.
\begin{defn}
We say that modulus consensus is established by the protocol~\eqref{eq.altaf} if the coincident limits exist
\[
\lim_{k\to\infty}|\xi_1(k)|=\ldots=\lim_{k\to\infty}|\xi_n(k)|\quad\text{for any $\xi(0)\in\r^n$}.
\]
\end{defn}

The absolute values $x_i(k)=|\xi_i(k)|$ obey the inequalities
\be
x_i(k+1)\le \sum_{j=1}^n|a_{ij}(k)|x_j(k)\quad\forall i,
\ee
and hence the vector $x(k)=(x_1(k),\ldots,x_n(k))^{\top}$ obeys~\eqref{eq.ineq} with $W(k)=|A(k)|$. If this recurrent inequality is consensus dichotomic, then modulus consensus in~\eqref{eq.altaf} is established. Theorems~\ref{thm.2} and \ref{thm.3} yield in in the following criterion.
\begin{thm}\label{thm.altaf}
Modulus consensus in~\eqref{eq.altaf} is established, if the sequence of matrices $W(k)=|A(k)|$ satisfies the conditions of Theorem~\ref{thm.2} or Theorem~\ref{thm.3}.
\end{thm}

In particular, if Assumption~\ref{ass.positive} holds, then modulus consensus is ensured by the repeated strong connectivity (Assumption~\ref{ass.uniform}), which can be relaxed to the infinite strong connectivity (Assumption~\ref{ass.inf}) if the network is bidirectional $w_{ij}(k)>0\Leftrightarrow w_{ji}(k)>0$. Theorem~\ref{thm.altaf} includes thus the results of Theorems~2.1 and~2.2 in~\cite{MengShiCao:16}. As discussed in Section~\ref{sec.main}, the condition from Theorem~\ref{thm.2} holds in many situations where Assumption~\ref{ass.positive} fails, e.g. $W(k)\equiv W$ may be
a constant primitive matrix with zero diagonal entries. Unlike consensus algorithms~\eqref{eq.eq}, where the gains $w_{ij}(k)$ are design parameters, the social influence (or ``social power'') of an individual over another one depends on many uncertain factors~\cite{FrenchRaven:1959}, and the uniform positivity of the non-zero gains $|a_{ij}(k)|$ may become a restrictive assumption.

In general, the assumptions of Theorems~\ref{thm.2} and~\ref{thm.3} do not guarantee the \emph{exponential} convergence rate to the equilibrium, which is provided by Assumptions~\ref{ass.positive} and~\ref{ass.uniform}~\cite{LiuChenBasar:15,LiuChamieBasarAcikmese:16}. In the case of exponential convergence, an additional criterion has been established in~\cite{LiuChenBasar:15,LiuChamieBasarAcikmese:16} (see also Theorem~2.3 in~\cite{MengShiCao:16}), allowing to distinguish between ``degenerate'' modulus consensus (asymptotic stability
of the linear system~\eqref{eq.altaf}) and \emph{polarization}. In the latter case, the agents split into two ``hostile camps'' $V_1\cup V_2=V=\{1,\ldots,n\}$,
and the opinions of agents from $V_i$ converge to $(-1)^iM$, where $M=M(\xi(0))\ne 0$ for almost all $\xi(0)$. If $V_1=\emptyset$ or $V_2=\emptyset$, then polarization reduces to usual consensus of opinions.

\subsection{Substochastic matrices and the Friedkin-Johnsen model}

A non-negative matrix $A=(a_{ij})$ is called \emph{substochastic} if $\sum_{j=1}^na_{ij}=1\,\forall i$.
We say that the $i$th row of $A$ is a \emph{deficiency} row of $A$ if the latter inequality is strict $\sum_ja_{ij}<1$.
Unlike a stochastic matrix, always having an eigenvalue at $1$, a substochastic square matrix is usually Schur stable $\rho(A)<1$. Theorem~\ref{thm.1} allows to give an elegant proof of the Schur stability
criterion for substochastic matrices~\cite{FrascaTempo:2013,Parsegov2017TAC}.
\begin{lem}\label{lem.substoch}
Let $\g=\g[A]$ be the graph of a substochastic square matrix $A$ and $I_d=\{i:\sum_ja_{ij}<1\}$ is the subset of its nodes, corresponding to the deficiency rows of $A$.
If any node $j$ either belongs to the set $I_d$, or $I_d$ is reachable from it in $\g$ via some walk, then $\rho(A)<1$.
\end{lem}
\begin{proof}
Consider the matrix $W=(w_{ij})$, defined by
\[
w_{ij}\dfb a_{ij}+\frac{1}{n}\left(1-\sum_la_{il}\right)\ge a_{ij}.
\]
Obviously, $W=(w_{ij})$ is stochastic and $w_{ij}>a_{ij}\ge 0\,\forall j$ when $i\in I_d$. Hence in the graph $\g[W]$ each node $i\in I_d$ is connected to any other node and to itself, and hence $\g[W]$ is aperiodic. The condition of Lemma~\ref{lem.substoch} implies that $\g[W]$ is also strongly connected. Choosing an arbitrary non-negative vector $x_0\ge 0$, the vectors $x(k)=A^kx_0$ are non-negative for any $k\ge 0$ and satisfy the inequality~\eqref{eq.ineq} with
$W(k)\equiv W$. Thanks to Theorem~\ref{thm.1}, $x(k)\to c\ones$, where $c\ge 0$. It remains to notice that $\ones$ is not an eigenvector of $A$ since $I_d(A)\ne\emptyset$, and hence $c=0$.
Thus $A^kx_0\to 0$ as $k\to\infty$ for any $x_0\ge 0$, which implies the Schur stability of $A$ since any vector $x_0$ is a difference of two non-negative vectors.
\end{proof}

Notice that Lemma~\ref{lem.substoch} implies the following well-known property of substochastic irreducible matrices~\cite{Meyer2000Book}: if $\g$ is strongly connected then $A$ is either stochastic or Schur stable. The condition from Lemma~\ref{lem.substoch} is not only sufficient but also necessary for the Schur stability~\cite{Parsegov2017TAC}.
Lemma~\ref{lem.substoch} implies the condition of opinion convergence in the \emph{Friedkin-Johnsen} model of opinion formation~\cite{FriedkinJohnsen:1999,Friedkin:2015,Parsegov2017TAC}
\be\label{eq.fj}
x(k)=\La W x(k)+(I-\La)u,\quad u=x(0).
\ee
Here $W$ is a stochastic matrix of influence weights, and $\La$ is a \emph{diagonal} matrix of the agents' \emph{susceptibilities} to the social influence~\cite{FriedkinJohnsen:1999}, $0\le\la_{ii}\le 1$. Without loss of generality, one may suppose that $\la_{ii}=0\Leftrightarrow w_{ii}=1$; in this case agent $i$ is \emph{stubborn} $x_i(k)\equiv x_i(0)$ (often it is assumed~\cite{FriedkinJohnsen:1999} that $\la_{ii}=1-w_{ii}$). Another extremal case is $\la_{ii}=1$, which means that agent $i$ ``forgets'' its initial opinion $u_i=x_i(0)$ and iterates the usual
procedure of opinion averaging $x_i(k+1)=\sum_{j}w_{ij}x_j(k)$. If $0<\la_{ii}<1$, then agent $i$ is ``partially stubborn'' or \emph{prejudiced}~\cite{ProTempo:2017-1,ProTempoCao16-2}: such an agent adopts the others' opinions, however it is ``attached'' to its initial opinion $x_i(0)$ and factors it into every opinion iteration.

If the substochastic matrix $\La W$ is Schur stable, then the opinion vector $x(k)$ in~\eqref{eq.fj}
converges to the equilibrium
\be\label{eq.fj-final}
x(k)\xrightarrow[k\to\infty]{}(I-\La W)^{-1}(I-\La)u.
\ee
By noticing that the graphs $\g[\La W]$ and $\g[W]$ differ only by the structure of self-loops (recall that $\la_{ii}>0$
unless $w_{ii}=1$ and $w_{ij}=0\,\forall j\ne i$),
Lemma~\ref{lem.substoch} implies the following.
\begin{cor}~\cite{Parsegov2017TAC}
The opinions~\eqref{eq.fj-final} converge if from each agent $i$ with $\la_{ii}=1$ there exists a
walk in $\g[W]$ to some agent $j$ with $\la_{jj}<1$, that is, each agent is either prejudiced or influenced (directly or indirectly) by a prejudiced agent.
\end{cor}

Using Theorems~\ref{thm.2} and~\ref{thm.3}, some stability criteria for the time-varying extension~\cite{ProTempoCao16-2} of the Friedkin-Johnsen model can be obtained that are beyond the scope of this paper.

\subsection{Constrained consensus}

In this subsection, we consider another application of the recurrent inequalities case, related to the problem of \emph{constrained} or ``optimal'' consensus that is closely related to distributed convex optimization~\cite{Nedic:10,ShiJohansson:2012,LinRen:14} and distributed algorithms, solving linear equations~\cite{LiuMorseNedicBasar:14,MouLiuMorse:15,YouSongTempo:16}.

\begin{figure}
\center
\includegraphics[height=2.5cm]{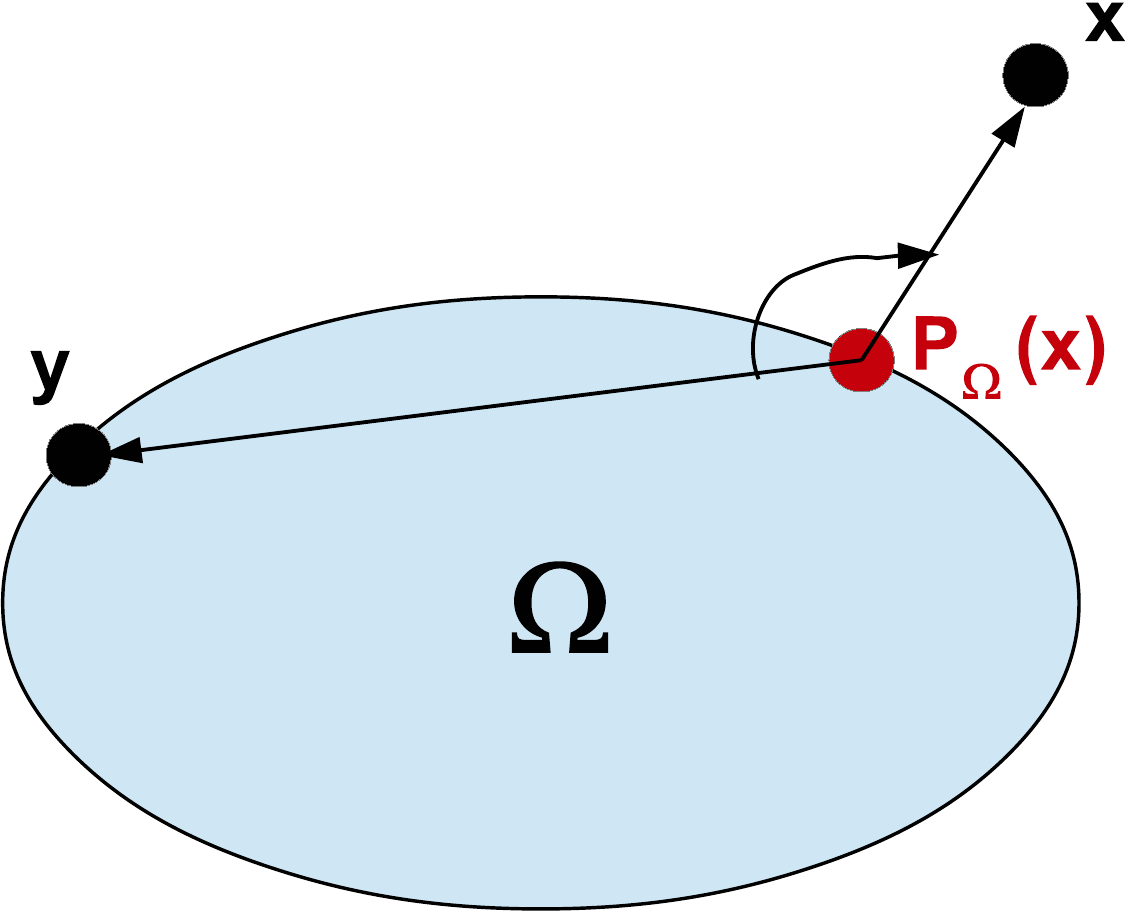}
\caption{The projection onto a closed convex set}\label{fig.proj}
\end{figure}

For any closed convex set $\Omega\subset\r^d$ and $x\in\r^d$ the \emph{projection} operator $P_{\Omega}:x\in\r^d\mapsto P_{\Omega}(x)\in\Omega$ can be defined, mapping a point to the closest
element of $\Omega$, i.e. $\|x-P_{\Omega}(x)\|=\min_{y\in\Omega}\|x-y\|$. This implies that $\measuredangle(y-P_{\Omega}(x),x-P_{\Omega}(x))\ge\pi/2$ (Fig.~\ref{fig.proj}) and
\be\label{eq.proj}
\|x-y\|^2\ge \|x-P_{\Omega}(x)\|^2+\|y-P_{\Omega}(x)\|^2\quad\forall y\in\Omega.
\ee
The distance $d_{\Omega}(x)\dfb\|x-P_{\Omega}(x)\|$ is a convex function.

Consider a group of $n$ discrete-time agents with the state vectors $\xi_i(k)\in\r^d$. Each agent is associated with a closed convex set $\Xi_i\subseteq\r^d$ (e.g., the set of minima of some convex function).
The agents' cooperative goal is to find some point $\xi_*\in \Xi\dfb \Xi_1\cap\ldots\cap \Xi_n$.
To solve this problem, various modifications of the protocol~\eqref{eq.eq} have been proposed. We consider the following three algorithms
\begin{gather}
\xi_i(k+1)=P_{\Xi_i}\left[\sum\nolimits_{j=1}^nw_{ij}(k)\xi_j(k)\right]\label{eq.nedic},\\
\xi_i(k+1)=P_{\Xi_i}\left[\sum\nolimits_{j=1}^nw_{ij}(k)P_{\Xi_j}(\xi_j(k))\right]\label{eq.morse},\\
\xi_i(k+1)=w_{ii}(k)P_{\Xi_i}(\xi_i(k))+\sum_{j\ne i}w_{ij}(k)\xi_j(k)\label{eq.tempo}.
\end{gather}
Here $W(k)=(w_{ij}(k))$ stands for the sequence of stochastic matrices.
The protocol~\eqref{eq.nedic} has been proposed in the influential paper~\cite{Nedic:10} (see also~\cite{LinRen:14}), dealing with distributed optimization problems.
The special cases of protocols~\eqref{eq.morse} and~\eqref{eq.tempo} naturally arise in distributed algorithms, solving linear equations, see respectively~\cite{LiuMorseNedicBasar:14,MouLiuMorse:15} and~\cite{YouSongTempo:16}; a randomized version of~\eqref{eq.tempo} has been also examined in~\cite{ShiJohansson:2012}.
\begin{thm}
Let the set $\Xi_i$ be closed and convex, and assume that $\Xi=\Xi_1\cap\ldots\cap\Xi_n\ne\emptyset$. Suppose that the matrices $W(k)$ satisfy Assumptions~\ref{ass.positive} and~\ref{ass.uniform}.
Then each of the protocols~\eqref{eq.nedic}-\eqref{eq.tempo} establishes \emph{constrained} consensus:
\be\label{eq.cons-cons}
\lim_{k\to\infty}x_1(k)=\ldots=\lim_{k\to\infty}x_n(k)\in\Xi.
\ee
\end{thm}
\begin{proof}
Due to the page limit, we give only an outline of the proof. By assumption, there exists some $\xi_0\in\Xi$. Denote $P_i(\cdot)\dfb P_{\Xi_i}(\cdot)$, $d_i(\cdot)\dfb d_{\Xi_i}(\cdot)$ and let
$\eta_i(k)\dfb\sum_{j}w_{ij}(k)\xi_j(k)$.
Under Assumptions~\ref{ass.positive} and~\ref{ass.uniform}, to prove the constrained consensus~\eqref{eq.cons-cons} it suffices to show~\cite{LinRen:14} that
\be\label{eq.e}
e_i(k)\dfb \xi_i(k+1)-\eta_i(k)\xrightarrow[k\to\infty]{} 0,\quad d_i(\xi_i(k))\xrightarrow[k\to\infty]{} 0.
\ee

Applying~\eqref{eq.proj} to $\Omega=\Xi_i$, $x=\xi$, $y=\xi_0\in\Xi_i$, one gets
\be\label{eq.proj1}
\|\xi-\xi_0\|^2\ge \|P_i(\xi)-\xi_0\|^2+d_i(\xi)^2\quad \forall\xi\in\r^d,
\ee
and therefore $\|\xi-\xi_0\|\ge \|P_i(\xi)-\xi_0\|$. Each protocol \eqref{eq.nedic}-\eqref{eq.tempo} thus implies the recurrent inequality~\eqref{eq.ineq}, where $x_i(k)\dfb \|\xi_i(k)-\xi_0\|\,\forall i$. For instance, the equation~\eqref{eq.nedic} entails that
\[
0\le x_i(k+1)\le \left\|\sum_{j=1}^nw_{ij}(k)\xi_j(k)-\xi_0\right\|\le \sum_{j=1}^nw_{ij}(k)x_j(k).
\]
Corollary~\ref{cor.1} implies the existence of the common limit $x_*=\lim_{k\to\infty}x_i(k)\ge 0$. We are now going to prove~\eqref{eq.e} for the protocol~\eqref{eq.nedic}.
The second statement in~\eqref{eq.e} is obvious since $d_i(\xi_i(k+1))\equiv 0$. Substituting $\xi=\eta_i(k)$ into~\eqref{eq.proj1},
\be\label{eq.tech3}
\begin{split}
\|e_i(k)\|^2\overset{\eqref{eq.nedic}}{=}d_i(\eta_i(k))^2\overset{\eqref{eq.proj1}}{\le} \|\eta_i(k)-\xi_0\|^2-x_i(k+1)^2\le\\
\le\sum\nolimits_jw_{ij}(k)x_j(k)-x_i(k+1)\xrightarrow[k\to\infty]{} 0.
\end{split}
\ee
To prove~\eqref{eq.e} for the protocol~\eqref{eq.tempo}, notice that
\be\label{eq.tech4}
\begin{split}
x_i(k+1)\overset{\eqref{eq.nedic}}{\le} w_{ii}(k)\|P_i(\xi_i(k))-\xi_0\|+\sum_{j\ne i}w_{ij}x_j(k)\\
\overset{\eqref{eq.proj1}}{\le}
 w_{ii}(k)\sqrt{x_i(k)^2-d_i(\xi_i(k))^2}+\sum_{j\ne i}w_{ij}x_j(k).
\end{split}
\ee
Recalling that $w_{ii}(k)\ge\delta$ and $x_i(k)\to x_*\forall i$, it can be shown that $d_i(\xi_i(k))\to 0$ and hence $\|e_i(k)\|=w_{ii}(k)d_i(\xi_i(k))\to 0$. The property~\eqref{eq.e} for the protocol~\eqref{eq.morse} is
proved similarly, combining the arguments from~\eqref{eq.tech3} and~\eqref{eq.tech4}.
\end{proof}

\section{Conclusions}

In this paper, we have examined a class of recurrent inequalities~\eqref{eq.ineq}, inspired by the analysis of ``modulus consensus'' in signed networks.
Under natural connectivity assumptions the inequality is shown to be \emph{consensus dichotomic}, that is, any of its solution is either unbounded
or converges to consensus. Besides signed networks, we illustrate the applications of this profound property to some problems of matrix theory and distributed optimization algorithms.

\bibliographystyle{IEEETran}
\bibliography{consensus}

\appendix
\section*{Proofs of Theorems~\ref{thm.1}-\ref{thm.3}}

Although the results of Theorems~\ref{thm.1}-\ref{thm.3} resemble usual criteria of convergence for consensus protocols~\eqref{eq.eq}, their proofs are based on different techniques,
which are inspired by the analysis of differential inequalities from~\cite{ProCao16-4,ProCao:2017}.
Given a solution $x(k)\in\r^n$ of~\eqref{eq.ineq}, let $j_1(k),\ldots,j_n(k)$ be the permutation of the indices $\{1,\ldots,n\}$, sorting the elements of $x(k)$ in the
\emph{ascending} order. In other words, the numbers $y_i(k)\dfb x_{j_1(k)}(k)$ satisfy the inequalities
\ben
\min_i x_i(k)=y_1(k)\le y_2(k)\le\ldots\le y_n(k)=\max_ix_i(k).
\een
We also introduce the sets $J_i(k)\dfb\{j_1(k),\ldots,j_i(k)\}$ and $J_i^c(k)\dfb\{1,\ldots,n\}\setminus J_i(k)$ (where $1\le i\le n$).
Recall that $\Phi(m,k)=(\vp_{ij}(m_s,k_s))=W(m-1)\ldots W(k)$ stands for the evolutionary matrix of the linear equation~\eqref{eq.eq}.

We start with the following simple proposition.
\begin{prop}\label{lem.tech}
For any $p\ge 0$, $q>p$ and $j,l=1,\ldots,n$ the inequality holds as follows
\ben
x_l(q)\le y_n(p)-\vp_{lj}(q,p)\left(y_n(p)-x_j(p)\right)\label{eq.tech1}.
\een
In particular, for the case where $q=p+1$ one has
\be
x_l(p+1)\le y_n(p)-w_{lj}(p)\left(y_n(p)-x_j(p)\right)\label{eq.tech2}.
\ee
\end{prop}
\begin{proof}
The proof is immediate from the inequalities
\[
\begin{split}
x_l(q)\le\sum_{s=1}^n\vp_{ls}(q,p)x_s(p)&=\vp_{lj}(q,p)x_j(p)+\\&+\sum_{s\ne j}\vp_{ls}(q,p)x_s(p)
\end{split}
\]
by noticing that $x_s(p)\le y_n(p)$ for any $s\ne j$.
\end{proof}

\begin{lem}\label{lem.tech1}
Suppose that for some $k\ge 0$ and $m>k$ one has $\Phi(m,k)\in S_{\ve}$. Then for any $i<n$ one has
\be\label{eq.tech}
y_{i+1}(m)\le (1-\ve)y_n(k)+\ve y_i(k).
\ee
\end{lem}
\begin{proof}
The definition of the set $J_i(k)$ and~\eqref{eq.tech1} imply that $x_j(m)\le (1-\ve)y_n(k)+\ve y_i(k)$ for any $j\in J_i(k)$ since $\vp_{jj}(m,k)\ge\ve$ and $x_j(k)\le y_i(k)$.
Since the graph $\g_{\ve}[\Phi(m,k)]$ is strongly connected,
there exist some $j\in J_i(k)$ and $l\in J_i^c(k)$ such that $\vp_{lj}(m,k)\ge\ve$, and thus $x_{l}(m)\le (1-\ve)y_n(k)+\ve y_i(k)$ due to~\eqref{eq.tech1}.
This entails~\eqref{eq.tech} since the set $J_i(k)\cup\{l\}$ contains $i+1$ elements.
\end{proof}

The statement of Lemma~\ref{lem.tech1} retains its validity, replacing the condition from Theorem~\ref{thm.2} by the assumptions of Theorem~\ref{thm.3}. However, in this situation $m=m(k,i)$ should be chosen in a different way and depends on both $i$ and $k$.
\begin{lem}\label{lem.tech2}
Let Assumptions~\ref{ass.positive} and~\ref{ass.inf} hold and the communication be bidirectional $w_{ij}(k)>0\Longleftrightarrow w_{ji}(k)>0$.  Then for all $k\ge 0$, $i<n$ there exists $m=m(i,k)>k$ such that
\be\label{eq.tech+}
y_{i+1}(m)\le (1-\delta)y_n(k)+\delta y_i(k),
\ee
where $\delta>0$ is the constant from Assumption~\ref{ass.positive}.
\end{lem}
\begin{proof}
For a fixed $k\ge 0$ and $i<n$, we denote for brevity $J=J_i(k)$ and $J^c=J_i^{c}(k)$. Assumption~\ref{ass.inf} implies the existence of $s\ge k$ such that $w_{jl}(s)>0$ (and thus $w_{lj}>0$) for some $j\in J$ and $l\in J^c$. Let $m-1$ stand for the \emph{minimum} of such $s$ (that is, $m\ge k+1$). Since $w_{jl}(s)=w_{lj}(s)=0$ for any $s=k,\ldots,m-2$, one has
$\vp_{jl}(m-1,k)=\vp_{lj}(m-1,k)=0$ for any pair $j\in J,l\in J^c$. Hence
\[
x_j(m-1)=\sum_{r\in J}\vp_{jr}(m-1,k)x_r(k)\le y_i(k).
\]
Applying~\eqref{eq.tech2} to $p=m-1$ and $j=l\in J$, one has $x_j(m)\le (1-\delta)y_n(m-1)+\delta x_j(m-1)\le (1-\delta)y_n(k)+\delta y_i(k)$.
At the same time, there exist $j\in J,l\in J^c$ such that $w_{lj}(m-1)\ge\delta$. In view of~\eqref{eq.tech2}, $x_l(m)\le (1-\delta)y_n(m-1)+\delta x_j(m-1) \le (1-\delta)y_n(k)+\delta y_i(k)$.
This entails~\eqref{eq.tech+} since the set $J_i(k)\cup\{l\}$ contains $i+1$ elements.
\end{proof}

\subsection{Proofs of Theorems~\ref{thm.2} and~\ref{thm.3}}

Consider a bounded solution $x(k)$ of~\eqref{eq.ineq} and its ordering $y(k)$.
The inequality~\eqref{eq.ineq} implies, obviously, that $y_n(k+1)\le y_n(k)$, and therefore there exists the limit $y_*=\lim_{k\to\infty}y_n(k)$.
Our goal is to show that $y_j(k)\xrightarrow[k\to\infty]{} y_*$ for any $j$, provided that the assumptions of either Theorem~\ref{thm.2} or Theorem~\ref{thm.3} are valid.
The proof is via induction on $j=n,n-1,\ldots,1$. For $j=n$ the statement holds. Suppose that $y_{j}(k)\to y_*$ for $j=i+1,\ldots,n$; we are now going to prove that $y_{i}(k)\to y_*$ as $k\to\infty$.
Since $y_i(k)\le y_n(k)$,  it suffices to show that $\varliminf\limits_{k\to\infty}y_i(k)\ge y_*$. Suppose, on the contrary, that
$\varliminf\limits_{k\to\infty}y_i(k)<y_*$, that is, there exist a sequence $k_s\xrightarrow[s\to\infty]{}\infty$ and $q>0$, such that $y_i(k_s)\xrightarrow[s\to\infty]{} y_*-q$.

As implied by Lemma~\ref{lem.tech1} (respectively, Lemma~\ref{lem.tech2}), under the assumptions of Theorem~\ref{thm.2} (respectively, Theorem~\ref{thm.3}),
a sequence $m_s>k_s$ and a constant $\ve>0$ exist such that
\[
y_{i+1}(m_s)\le\ve y_i(k_s)+(1-\ve)y_n(k_s).
\]
Passing to the limit as $s\to\infty$, one arrives at
\[
\begin{split}
y_*=\lim_{s\to\infty}y_{i+1}(m_s)\le (1-\ve)y_*+\ve (y_*-q)=y_*-\ve q,
\end{split}
\]
which is a contradiction. Thus $y_i(k)\to y_*$ as $k\to\infty$, which proves the induction step.
Therefore, the solution converges to a consensus equilibrium $x(k)\xrightarrow[k\to\infty]{} y_*\ones_n$.
\hfill $\blacksquare$

\subsection{Proof of Theorem~\ref{thm.1}}

The sufficiency part is immediate from Remark~\ref{rem.2} and Theorem~\ref{thm.2}. Indeed, if the graph $\g[W]$ is strongly connected and aperiodic, then $W^d$ is a strictly positive matrix for some $d$, so the condition of Theorem~\ref{thm.2} holds: $\Phi(k+d,k)=W^d\in S_{\ve}$ for some $\ve>0$. Hence the inequality~\eqref{eq.ineq} is consensus dichotomic.
If the graph $\g[W]$ is constituted by $s>1$ isolated and aperiodic strongly connected components, then~\eqref{eq.ineq} is dichotomic, reducing to $s$ \emph{independent} consensus dichotomic inequalities of lower dimensions.

To prove necessity, consider a dichotomic inequality~\eqref{eq.ineq} with $W(k)\equiv W$ and let $w_{ij}>0$, that is, $i$ is connected to $j$ in the graph $\g[W]$.
Let the set $J$ include node $j$ and all nodes that are reachable from $j$ by walks.
We are going to show that $i\in J$, that is, $i$ and $j$ belong to the same strong component. Suppose, on the contrary, that $i\not\in J$ and let
\[
x_r(k)=
\begin{cases}
(-1)^k,&r=i,\\
M,&r\in J,\\
-1,&r\not\in J\cup\{i\},
\end{cases}\quad r=1,\ldots,n.
\]
Here $M$ is chosen sufficiently large so that $(M+1)w_{ij}>2$.
It can be easily shown that the vector $x(k)$ is a solution to~\eqref{eq.ineq}. Indeed, for any $r\in J$ and $q\not\in J$ one obviously has $w_{rq}=0$ (otherwise, $q$ would be reachable from $j$ via $l$). Therefore, $M=x_r(k)=\sum_{q=1}^n w_{rq}x_q(k)=\sum_{q\in J}w_{rq}x_q(k)$. For any $r\not\in J\cup\{i\}$ we have $x_r(k)=\min_qx_q(k)\le \sum_{q=1}^n w_{rq}x_q(k)$.
Finally, $x_i(k)\le 1\le Mw_{ij}-(1-w_{ij})=Mw_{ij}-\sum_{q\ne j}w_{iq}\le w_{ij}x_j(k)+\sum_{q\ne j}w_{iq}x_q(k)=\sum_{q=1}^nw_{iq}x_q(k)$.
Since $x(k)$ is bounded yet does not converge, one arrives at the contradiction with the assumption of dichotomy.

Hence for the dichotomy it is necessary that any arc connects nodes from the same strong components, whereas two different components have no arcs between them. Notice that the dichotomy (respectively, consensus dichotomy) of the inequality~\eqref{eq.ineq} implies that any solution of the equation~\eqref{eq.eq} converges to an equilibrium (respectively, to a consensus equilibrium). Applying the standard convergence and consensus criteria for the static consensus protocol~\eqref{eq.eq} (see e.g.~\cite{ProTempo:2017-1}), one shows that dichotomy is possible only when all strong components of $\g[W]$ are aperiodic, whereas consensus dichotomy implies that the graph is strongly connected and aperiodic, which ends the proof. 

\end{document}